\documentstyle[aps,preprint,tighten,floats,epsf,rotate]{revtex}

\newcommand{\bea}{\begin{eqnarray}}
\newcommand{\eea}{\end{eqnarray}}
\newcommand{\be}{\begin{equation}}
\newcommand{\ee}{\end{equation}}

\begin{document}

\draft
%%%%%%%%%%%%%%%%%%%%%%%%%%%%%%%%%%%%%%%%%%%%%%%%%%%%%%%%%%%%%%%%%%%%%%%
\preprint{\vbox{\it
                        \null\hfill\rm   SU-4252-757 \\ \null\hfill\rm
BI-TP 2002/09\\ \null\hfill\rm 22th April 2002}\\\\}
%%%%%%%%%%%%%%%%%%%%%%%%%%%%%%%%%%%%%%%%%%%%%%%%%%%%%%%%%%%%%%%%%%%%%%%

\title{ Non-abelian Topological Strings and Metastable States in 
Linear Sigma Model}

\author{A. P. Balachandran\footnote{bal@phy.syr.edu}}
\address{Physics Department, Syracuse University, Syracuse,
New York 13244-1130}
\author{S. Digal\footnote{digal@physik.uni-bielefeld.de}}
\address{Fakult\"{a}t f\"{u}r Physik,
Universit\"{a}t Bielefeld, D-33615, Bielefeld, Germany}

\maketitle
\widetext

\begin{abstract}

Non-abelian (NA) topological string defects exist in QCD with the spontaneous
chiral symmetry breaking $U(N_f)_L \times U(N_f)_R \to U(N_f)_V$. Anomaly 
effects lead to domain walls connected to these strings. In the framework of 
linear sigma model we find the configuration of these defects. Also we show 
that in this model metastable $CP$ violating vacuua exist. Strings as 
well as extended regions of metastable vacuua may form during the chiral 
transition which could have interesting effects.

\end{abstract}


\section{\bf Introduction}

QCD with $N_f$ flavors of massless quarks possesses the chiral symmetry 
$U(N_f)_L \times U(N_f)_R$. The generators of its diagonal subgroup
$U(N_f)_V$ are parity invariant while the parity-odd generators give the
axial transformations $U(N_f)_A$. (The latter do not form a group.) For low 
energy and temperatures the axial vector part of this symmetry suffers 
spontaneous symmetry breakdown (SSB), this is indicated by 
$U(N_f)_V \times U(N_f)_A \longrightarrow U(N_f)_V$ \cite{vafa}. 
This creates $N_f^2$ massless pseudoscalar Goldstone bosons known as 
mesons. However in the real world, chiral symmetry is explicitly broken
because of non-zero quark masses. This makes the pseudoscalar mesons massive. 
Even in the chiral limit, anomaly effects break the $SU(N_f)_V$-invariant
axial $U(1)$ subgroup $U(1)_A \subset U(N_f)_A$ 
to $Z_{N_f}$ \cite{hooft} making the $\eta^\prime$ meson heavier compared 
to other mesons. Finite temperature lattice QCD studies show that
chiral symmetry (but for $U(1)_A$ of course which is only effectively restored)
is restored at temperatures, $T$ larger than a critical value $T_\chi$. Such 
chirally symmetric phase of matter is believed to have existed in the early 
Universe. The matter in the initial stages of the fire ball produced in heavy 
ion collisions is likely to be in the chirally symmetric phase. To see 
such a phase transition in laboratory is one of the motivations of heavy-ion 
collision experiments. 

The chirally symmetric and broken phases of QCD are characterized by an 
order parameter (OP) which is the quark-antiquark condensate known as the 
chiral condensate $\Phi$. It is an $N_f \times N_f$ matrix at each $x$.
In the chiral symmetry limit (with no anomaly), all possible values of $\Phi$ 
constitute the order parameter space (OPS). In the broken symmetry phase 
instead, it is diffeomorphic to $U(N_f)$. $U(N_f)$ as the OPS allows for the 
existence of ``abelian'' as well as ``non-abelian'' topological string 
defects. Brandenberger et al.\cite{berger} first studied the formation of 
``abelian'' strings due to SSB of the $U(1)_A$ part of chiral symmetry. 
In \cite{berger} it was argued that these string defects will exist as long 
as the anomaly effects are small and will decay as the anomaly becomes 
substantial. In a previous work we have considered the effects of anomaly on 
abelian strings \cite{bals}. We have argued that abelian strings will exist 
in the presence of anomaly with the typical structure of $N_f$ domain walls 
connected to the string. Now, a closer look at the full OPS=$U(N_f)$ shows 
that an additional class of topological defects, known as non-abelian (NA) 
strings can also exist, as the arguments in ref.\cite{bal} for example show. 
Note that here we use the phrase "non-abelian strings" in same way as 
the phrase "non-abelian monopoles" \cite{bal,nab} are used and not in the sense 
that $\pi_1(OPS)$ is non-abelian.
Here we find and discuss numerical solutions for these NA strings and note 
the important differences between the NA and abelian strings. Also we show 
that in the linear sigma model we employ, metastable vacua exist at zero 
temperature. Such metastable states have been found to exist in the
non-linear sigma model \cite{met} before, but for temperatures close to the 
deconfinement and/or chiral phase transition.  

In the next section we briefly discuss how the non-abelian string defects 
arise from the topological structure of the OPS=$U(N_f)$ during the SSB of 
chiral symmetry. In section III we find the numerical profiles for the NA 
strings using the linear sigma model Lagrangian \cite{jonathan}. In sec. IV 
we show existence of metastable CP violating states in the above model.
In section V we conclude with some speculative remarks on the formation of 
these string defects and metastable states during and/or after the chiral 
phase transition.

\section{\bf Non-abelian strings in QCD}

Topological defects are usually formed in phase transitions associated with
SSB with  non-trivial OPS. The type of defects formed depends on the dimension 
$n$ of physical space and the topology of the OPS. When the OPS after SSB 
is discrete, the topological defects are domain walls in 3-dimensional (3-D)
physical space. When the OPS after SSB is a circle the defects are vortices in 
2-D and strings in 3-D physical spaces. Monopoles arise in 3-D when the OPS
after SSB is a 2-sphere ($S^2$). For an introduction to and review of 
topological defects, see \cite{vilenkin}. For our discussion we need to know only about the string
defects. In the following we discuss briefly about how string defects arise. 

Topological string defects arise when there exist non-trivial loops in the
OPS, for example when it is a circle. A loop going around the circle $k$ times
cannot be smoothly shrunk to a point or cannot be smoothly deformed to 
another loop going $l \ne k$ times around the circle. So there exist 
non-trivial loops, characterized by windings, on the circle. There is a 
close connection between the non-trivial loops and the string defects in 
physical space, see Fig.1. A string defect of winding $k$  corresponds to 
a non-trivial loop winding $k$ times around a circle, not contractible to 
a point in the OPS, as a loop $\bf{l}$ around the string is traversed once 
in physical space. In Fig. 1 we show an example of winding number 1 
string defect for the case of a complex scalar field $\Psi$ with real 
and imaginary parts $\Psi_1$ and $\Psi_2$ when SSB is of $U(1)$. The figure
on the left of Fig.1 shows the typical effective potential for such a case. 
The minima $\bf{\it{M}}$ of the potential is a circle with radius $|\Psi|$. 
The configuration $\Psi$ of the string on the loop $\bf{l}$ (on $xy$-plane)
around the string core is shown on the right. Magnitude of $\Psi$ is 
represented by the length of the vectors and the phase by the angle $\theta$ 
these vectors make with the positive $x$-axis. As one goes around $\bf{l}$ 
once $\Psi$ on it traces a complete winding on $\bf{\it{M}}$. However, the 
configuration of $\Psi$ could have been such that it winds $k$-times (for a 
winding number  $k$ string) around $\bf{\it{M}}$ for a single loop on $\bf{l}$.

%%%%%%%%%%%%%%%%%%%%%%%%%%%%%%%%%%%%%%%%%%%%%%%%%%%%%%%%%%%%%%%%
\begin{figure}[ht]
\begin{center}
%\vskip -1 in
\leavevmode
\epsfysize=7truecm \vbox{\epsfbox{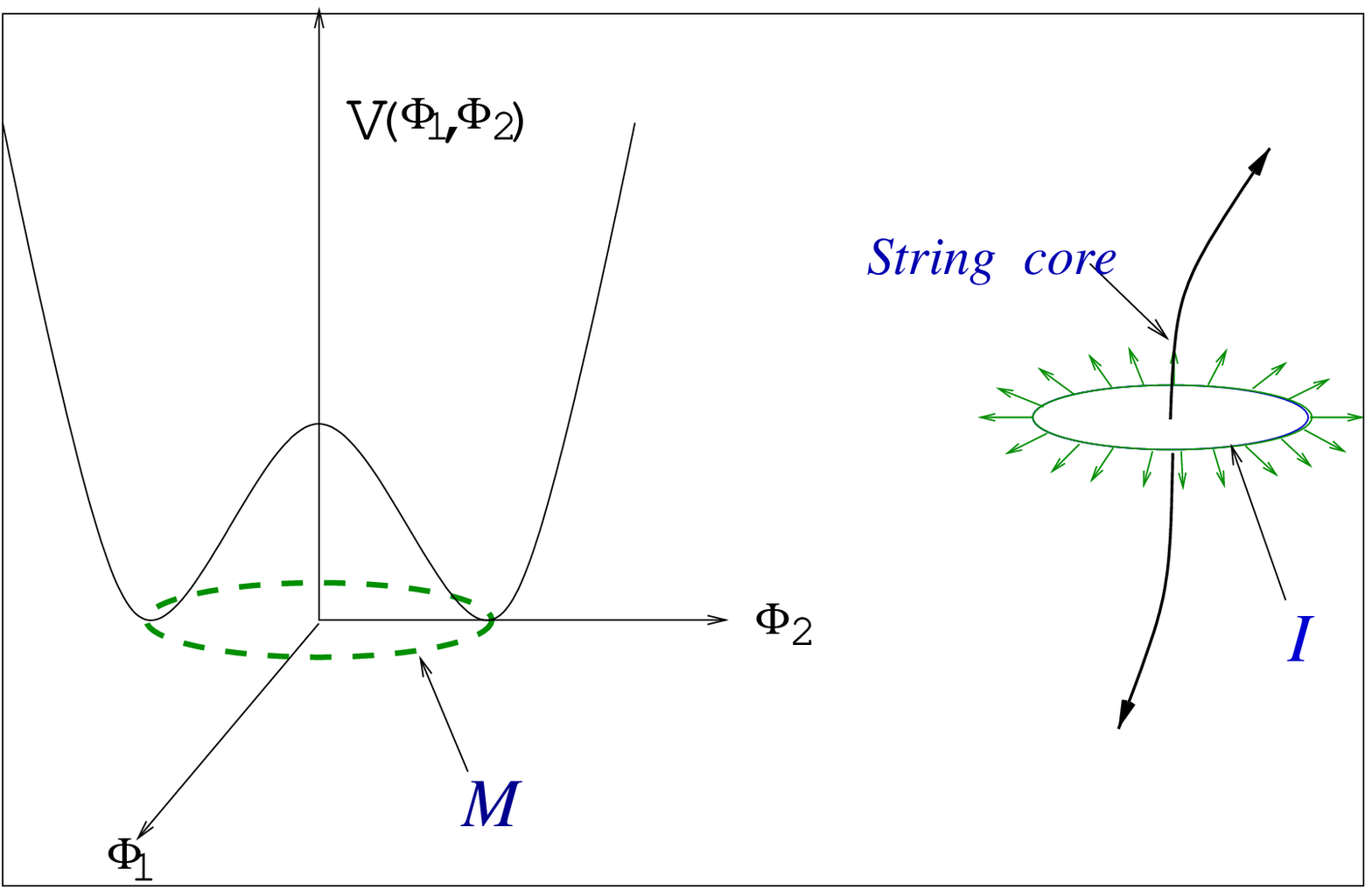}}
%\vskip -1in
\end{center}
\caption{On the left is the effective potential for SSB of $U(1)$. The
vacua are represented by the dashed circle. On the right is shown the 
configuration of the complex scalar field on the loop ${\it{\Large{\bf l}}}$ 
around the string core.}
\label{Fig.1}
\end{figure}
%%%%%%%%%%%%%%%%%%%%%%%%%%%%%%%%%%%%%%%%%%%%%%%%%%%%%%%%%%%%%%%%%%

The stability of the string defect is purely due to topological reasons. 
The total variation $\Delta\theta$ of $\theta$ around $\bf{l}$ can have only 
discrete values, namely integral multiples $2\pi k$ of $2\pi$. So by smooth 
changes of $\Psi$ on $\bf{l}$, one can not change $\Delta \theta$. 
Also by smoothly shrinking or enlarging $\bf{l}$, such that it does not cross
$\Psi=0$ when $\theta$ becomes undefined, one cannot change $\Delta \theta$ 
if the underlying $\Psi$ is continuous. On the other hand if $\bf{l}$ can be 
shrunk to a point ( keeping $\Psi$ continuous and non-zero except perhaps at
that point, and $\Delta \theta$ fixed in the process), then either $k=0$ or 
$\Psi$ must vanish at that point as it is continuous. By definition, 
$k \ne 0$ when there is a string (defect). The point where $\Psi=0$ when 
$k \ne 0$ is the core of the string. Strings in 3-D are either infinitely 
long or exist in loops. As strings have non-zero string tension, a string 
loop is unstable to shrinking.

It is important to note that string defects arise because one can have
loops with non-trivial winding on $\bf{\it{M}}$ which is a circle in the
above example. Such non-trivial loops exist in $U(N_f)$ which is the OPS after 
SSB of chiral symmetry. Thus let $T_a$ (a $\epsilon \{ 1, N_f^2-1\}$) be the
$SU(N_f)$ Lie algebra generators (the analogues of the Gell-mann matrices 
$\lambda_a$) and 

\begin{equation}
T_0= \sqrt{2 \over N_f}\bf{1}.
\end{equation}

Noting that

\begin{equation}
T_{N_f^2-1} =\left({2 \over N_f(N_f-1)} \right)^{1 \over 2} 
\left(\begin{array}{cccccc} 1 & & & & &   \\
  & 1 & & & &   \\
  &   &. & & &   \\
  & & &.& &   \\
  & & & &1& \\
  & & & & & 1-N_f  
\end{array}\right),
\end{equation}

\noindent a particular non-abelian string configuration is 

\begin{equation}
M(\theta)= e^{i {\theta \over N_f} \left[ N_f(N_f-1) \over 2 \right]^{1 
\over 2} T_{N_f^2-1}} e^{-i {\theta \over N_f} \left[ N_f 
\over 2 \right]^{1 \over 2} T_0}
\end{equation}
\noindent where $\theta$ is the angle on $\bf{\large{l}}$.
The first factor is an open curve in $SU(N_f)$ from $\bf{1}$ to its 
center $e^{i{2\pi \over N_f}} \bf{1}$. The second is an open curve in the 
center $U(1)$ of $U(N_f)$ from $\bf{1}$ to $e^{-i{2\pi \over N_f}} \bf{1}$.
On taking their product as in the above equation, the values at 
$\theta=2\pi$ multiply to $\bf{1}$ giving a closed string. The powers
$M(\theta)^N$ ($N \in Z \backslash \{0\}$) and their (continuous) 
deformations produce all possible strings, the deformations not changing their 
topological types. The beginning and end points of the non-abelian and abelian
factors remain fixed during these deformations.

Note that in $M(\theta)^{nN_f}$ ($n \in Z \backslash \{0\}$), the non-abelian
factor

\begin{equation}
e^{i n\theta \left[N_f(N_f-1) \over 2 \right]^{1 \over 2} T_{N_f^2-1}}
\end{equation}

\noindent is a closed loop in $SU(N_f)$. Such loops can be deformed to a 
point as the first homotopy group of $SU(N_f)$, $\pi_1(SU(N_f))$ is $0$. Thus 
$M(\theta)^{nN_f}$ gives the abelian strings. In other words, the elementary
strings are given by $M(\theta)$ and $M(\theta)^{-1}$. Note that by 
construction $M(\theta)^n$ gives a loop winding $n$ times. Consider the loop
given by 

\begin{equation}
M ^\prime (\theta)= e^{i {\theta \over N_f} \left[ N_f(N_f-1) \over 2 
\right]^{1 \over 2} T_{N_f^2-1}} e^{i {\theta (N_f - 1) \over N_f} \left[ N_f
\over 2 \right]^{1 \over 2} T_0},
\end{equation}

\noindent for which the open curve in $SU(N_f)$ remains the same as that
of $M(\theta)$, but one traverses the two end points in $U(1)$ in an 
anti-clockwise direction. This loop is actually homotopic to the loop given by
$M(\theta)^{1-N_f}$. That is to say, there is a string loop homotopic to 
that of $M(\theta)^{1-N_f}$ which gives the string configuration of (4).

In the above we have considered SSB of chiral symmetry. However in the
real world this symmetry is explicitly broken due to quark masses, and
also the $U(1)_A$ part of $U(N_f)_A$ is broken down to $Z_{N_f}$ due to axial
anomaly and instanton effects.The effect of anomaly changes the usual 
cylindrical structure of the string creating domain walls connecting to the 
string core. In the case of abelian strings, for $N_f=3$, the string gets 
connected to three domain walls \cite{bals}. In the case of elementary NA 
strings the number of domain walls can be either one or two as we explain in 
the next section. The effect of quark masses makes the string configurations 
non-static. However the string defects are stable against small fluctuations 
and can only decay due to annihilation with other defects with opposite 
windings. 
 
In the next section we will numerically find the configuration of the 
elementary non-abelian strings for the case of $N_f = 3$. For our calculations 
we use the linear sigma model Lagrangian considered in ref.\cite{jonathan}. 
Numerically it is easier to find the static configurations. Because of this 
we work in the chiral limit which allows for static solutions.

\section{Configurations of the non-abelian string}

To find the field configurations for the non-abelian strings we consider
the following $U(3)_L \times U(3)_R$ linear sigma model for $3$ quark 
flavors \cite{jonathan,rose},

\bea \label{L}
{\cal L}(\Phi) &=&
{\rm Tr}  \left( \partial_{\mu} \Phi^{\dagger}
\partial^{\mu} \Phi
-  m^2 \, \Phi^{\dagger}
\Phi \right) -
\lambda_{1} \left[ {\rm Tr}  \left( \Phi^{\dagger}
 \Phi  \right) \right]^{2} -
\lambda_{2} {\rm Tr}  \left( \Phi^{\dagger}
 \Phi  \right)^{2} \nonumber \\
&+& c \left[ {\rm Det} \left( \Phi \right) +
{\rm Det}  \left( \Phi^{\dagger} \right) \right]
+ {\rm Tr} \left[H  (\Phi + \Phi^{\dagger})\right] \,\, ,
\eea

\noindent where we take $c>0$ for specificity. $\Phi$ is a complex 
$3 \times 3$ matrix of fields of the scalar and pseudoscalar mesons:

\be
\Phi =T_{a} \, \phi_{a} =   T_{a} \, (\sigma_{a} +
        i \pi_{a}). \label{defphi}
\ee

%\section{\bf Symmetries of the Lagrangian}

The field $\Phi$ transforms under the chiral transformation 
$U(3)_L \times U(3)_R$ as 

\be \label{trans}
\Phi \longrightarrow U_{\ell} \, \Phi \, U_{r}^{\dagger} \,\,\, ,
\,\,\,\, U_{\ell,r} \equiv
\exp\left(i \, \omega_{\ell,r}^{a} \, T_{a}\right) \,\, .
\ee

\noindent One can rewrite the infinitesimal left-right transformations in 
terms of  vector-axial vector transformations with parameters
$w^a_{V,A}=(w^a_l\pm w^a_r)/2$. 
The infinitesimal form of the  $U(3)_L \times U(3)_R$ symmetry 
transformation (\ref{trans}) is

\be
\Phi \longrightarrow \Phi  +
        i \, \omega_{V}^{a} \,  \left[ T_{a}, \Phi \right] + i\,
        \omega_{A}^{a} \, \left\{ T_{a}, \Phi\right\} \,\, .
\ee

\noindent $\Phi$ is a singlet under the $U(1)_V$ transformations associated
with $exp(iw^0_V T^0)$. In QCD, they give rise to a conserved charge 
identified with the baryon number.

In the above model the determinant term takes into account the instanton
effect which explicitly breaks the $U(1)_A$ symmetry \cite{rose}. It is not 
clear whether such a term can be justified for high temperatures, because 
vanishing $\Phi$ makes the instanton effects to disappear \cite{pw}, though 
lattice results show that $U(1)_A$ is effectively restored at slightly higher 
temperatures \cite{alles} than the temperature at which $SU(N)_A$ symmetry
is restored. The last term where 
($H = T_a h_a$, $h_a =$ constants) is due to non-zero quark masses. When 
$c = H = 0$ and $\lambda_1 > 0$, and $\lambda_2 > 0$, the Lagrangian has a 
global symmetry $U(N_f)_V \times U(N_f)_A$ for $m^2 > 0$. For $m^2 <0$ the 
axial symmetry $U(N_f)_A$ is spontaneously broken to identity and the OPS is 
$U(N_f)$. This results in $N_f^2$ Goldstone bosons. For $N_f=3$ these 
Goldstone bosons are the $\pi$'s, $K$'s, $\eta$ and $\eta^\prime$. However 
when just $c \ne 0$, the $U(1)_A$ is further broken to $Z_3$ by the axial 
anomaly. $U(3)_A$ symmetry is, in addition, explicitly broken by non-zero 
quark masses. Also $SU(3)_V$ is explicitly broken due to the difference in 
$u,d$, and $s$ quark masses giving rise to splitting in the meson masses.

%\section{\bf Order parameter space}

In order to find the static non-abelian string configurations we take $H$
to be zero. Even when $H=0$, not all the non-abelian
strings are static in the presence of anomaly just like the abelian string. 
To illustrate this we consider the NA string corresponding to the non-trivial 
loop $M(\theta)$ [with $N_f=3$] on $U(3)$. Scaling $M(\theta)$ by the vacuum
$\Phi_0 = {\sigma_0 \over \sqrt{6}} \bf{1}$, we get the loop in the space of
fields $\Phi$ to be  

\begin{equation}
\Phi(\theta)= {\sigma_0 \over \sqrt{6}} M(\theta) = {1 \over \sqrt{6}}
\left(\begin{array}{ccc} \sigma_0 & 0 & 0 \\
0 & \sigma_0 & 0\\
0 & 0 & \sigma_0 e^{-i\theta} \end{array} \right).
\end{equation}

As only the last diagonal component of $\Phi$ varies with $\theta$, the
anomaly contribution to the effective Lagrangian from such field variation 
is 

\begin{equation}
{c \over 3\sqrt{6}}\sigma_0^3cos\theta.
\end{equation}

\noindent For such a term, the ground  state prefers the unique value 
$cos\theta=1$. Because of this, the $\theta$ distribution around the string 
will not be symmetric. In most of the region around the string, $\theta$ will 
take the value $0$ which minimizes the potential energy term (10). Non-zero 
$\theta$ values will be confined to a wall-like region. Because of non-zero 
surface tension of the wall, motion of the string towards the wall decreases 
the energy of the configuration. So one will not have static solutions in this
case. 

On the other hand, the loop corresponding to $M(\theta)^n$ 
in OPS gives rise to a string configuration with $|n|$ domain walls. This
configuration is static as the motion of the string along any domain wall
does not decrease the energy of the configuration. To find out the details
of one such NA string configuration, we consider $N_f=3$ and $n=-2$. Here 
$M(\theta)^{-2}$ gives a loop homotopic to the loop of $M^\prime(\theta)$. 
For $N_f=3$, $M^\prime(\theta)$ is given by $exp (i \theta T_s)$ where $T_s$ 
is

\begin{equation}
T_s = \left(\begin{array}{ccc} 1 & 0 & 0 \\
0 & 1 & 0\\
0 & 0 & 0 \end{array} \right).
\end{equation}

\noindent The corresponding loop in the space of fields 
$\Phi$ is from

\begin{equation}
\Phi(\theta)= {\sigma_0 \over \sqrt{6}} M^\prime(\theta) = {1 \over \sqrt{6}}
\left(\begin{array}{ccc} \sigma_0 e^{i\theta}& 0 & 0 \\
0 & \sigma_0 e^{i\theta}& 0\\
0 & 0 & \sigma_0 \end{array} \right).
\end{equation}

Writing $ \sqrt{3 \over 2} ({\phi_1 + i\phi_2})$ for
$\sigma_0 e^{i\theta}$, the effective Lagrangian for such a $\Phi$ 
deduced from (6), with $\lambda_1 = \lambda$, $\lambda_2=0$ and $H=0$, 
is given by

\bea \label{Lr}
{\cal L} &=&{1 \over 2}
\partial_{\mu} \phi_1\partial^{\mu} \phi_1
+ {1 \over 2}\partial_{\mu} \phi_2 \partial^{\mu} \phi_2
- {\cal V}(\phi_1,\phi_2),
\eea

\noindent where

\bea
{\cal V}(\phi_1,\phi_2)&=&
 {m^2 + {\lambda \sigma_0^2 \over 3} \over 2} \, 
\left (\phi_1^2 + \phi_2^2 \right )  +
{\lambda \over 4}\left (\phi_1^2 + \phi_2^2 \right )^2 \nonumber \\
&-& { c\sigma_0 \over 2 \sqrt{6}} \left ( \phi_1^2 - \phi_2^2 \right )
+ {m^2 \over 6}\sigma_0^2 + {\lambda\over 36}\sigma_0^4.
\eea

\noindent The values of the parameters in the above equation are 
$\lambda = 53.5322$, $c=6685.19 MeV$ and $m^2 = -(591.256 MeV)^2$. For these 
values of the parameters we get $\sigma_0 = 110.23 MeV$, 
the sigma meson mass $m_\sigma = 1000 MeV$, and the $\eta^\prime$ mass 
$m_{\eta^\prime} = 950 MeV$. Note that the anomaly term in 
${\cal V}$ expressed in $\theta$ is $-{c\sigma_0/ 2\sqrt{6}} 
cos(2\theta)$ for which $\theta=0$ and $\theta=\pi$ are degenerate. This 
leads to a string configuration attached to 2 domain walls. In the following 
we will find the approximate numerical profile of the string associated with 
$M^\prime (\theta)$.

%Including the case of abelian string, in the linear sigma model with anomaly, 
%there are strings connected by one, two, or three domain walls.
 
Considering the string along the $z$-direction, $\phi_i$ for the static 
string satisfy the following field equations:

\bea \label{fe}
\nabla^2\phi_1=
{\partial{{\cal V}(\phi_1,\phi_2)} \over \partial {\phi_1}},\nonumber \\
\nabla^2\phi_2={\partial{{\cal V}(\phi_1,\phi_2)} \over \partial {\phi_2}},
\eea

\noindent where $\nabla^2 = {\partial^2 \over \partial x^2} +
{\partial^2 \over \partial y^2}$. Since in the presence of anomaly, the string 
profile is non-cylindrical, one has to find the solution by an energy 
minimization technique. For this we consider an initial cylindrical 
configuration with $\dot{\phi_i}=0$ at $t=0$
and evolve it with the field equations derived from Eq.(6) with a 
dissipative term:

%%%%%%%%%%%%%%%%%%%%%%%%%%%%%%%%%%%%%%%%%%%%%%%%%%%%%%%%%%%%%%%%
\begin{figure}[ht]
\begin{center}
%\vskip -1 in
\leavevmode
{\epsfysize=7.0truecm \vbox{\epsfbox{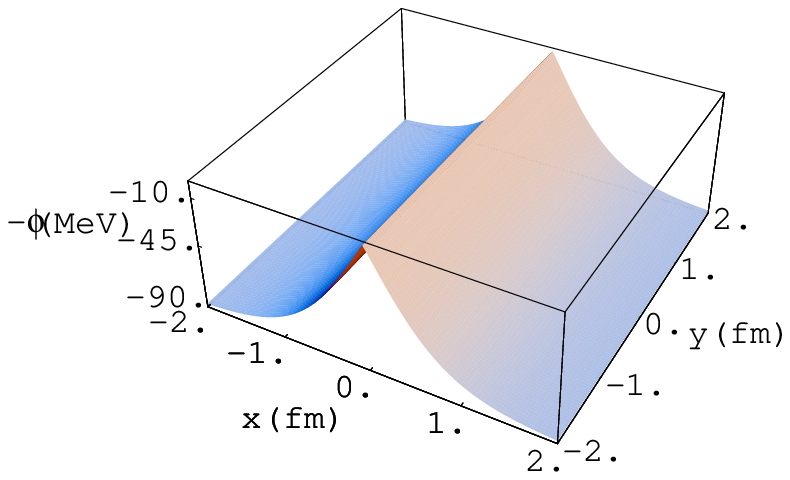}}
\epsfysize=7.0truecm \vbox{\epsfbox{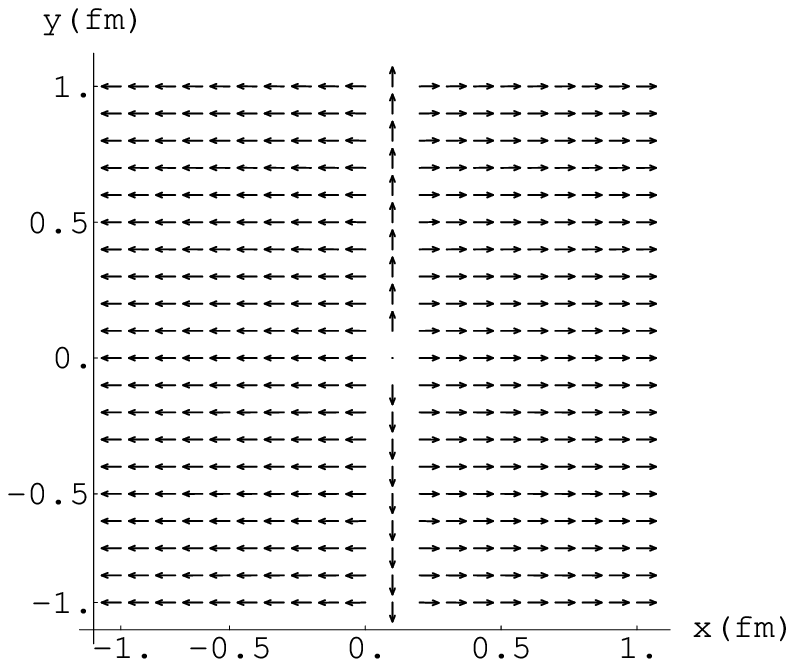}}}
%\vskip -1in
\end{center}
\caption{Configuration of the string for $c=6685.19 MeV$ and $H=0$.
Figure on the left gives $-\sqrt{\phi_1^2+\phi_2^2} \equiv -\phi$
and figure on the right gives the vector field plot of the full field 
$\phi_1(\vec{r})+i\phi_2(\vec{r})$.
}
\label{Fig.2}
\end{figure}
%%%%%%%%%%%%%%%%%%%%%%%%%%%%%%%%%%%%%%%%%%%%%%%%%%%%%%%%%%%%%%%%%%

\bea \label{fg}
{d^2\phi_1 \over dt^2} + \alpha {d\phi_1 \over dt} - \nabla^2\phi_1=
-{\partial{{\cal V}(\phi_1,\phi_2)} \over \partial {\phi_1}}\nonumber \\
{d^2\phi_2 \over dt^2} + \alpha {d\phi_2 \over dt} - \nabla^2\phi_2=
-{\partial{{\cal V}(\phi_1,\phi_2)} \over \partial {\phi_2}}.
\eea

\noindent Here $\alpha$ is the dissipation coefficient. The dissipation
term has been used to converge the initial configuration (the circular 
configuration, which is not the correct configuration in the presence of the 
anomaly) to the approximate configuration of the string connected to domain 
walls. Once the evolution is done for a sufficiently long time, the 
configuration does not evolve with time even after the dissipation is 
switched off, which suggests that the configuration is more or less static 
and so an approximate solution. On evolving the initial cylindrically 
symmetric configuration with the above equations, the domain walls connected 
to the string develop. In the numerical work,
we considered the $\eta^\prime$ mass to be the experimental value
$m_{\eta^\prime} \sim 950 MeV$ and set other pseudoscalar meson masses
to zero. In FIG.2 we show the configuration of the string with the domain
walls joined to it. In the left figure we plot
$-\sqrt{\phi_1^2+\phi_2^2}$, and in the right plot the vector field  
of $\Phi({\bf \vec{r}})=(\phi_1(\vec{r}),\phi_2(\vec{r}))$. The
magnitude of the field is proportional to the length of the vectors and the 
angle these vectors make with +ve $x$-axis is the phase. Clearly there is no 
symmetric distribution of the phase around the string. In the core of the 
string the full chiral symmetry associated with $T_s$ is restored.
That is because in the string core, as we explain below, $\Phi$ becomes
$diag(0,0,constant)$ which is invariant under transformations generated by
$T_s$. Although the field configuration has the structure of domain 
wall, the symmetry is restored only in the core of the string. It is not 
clear if it is the string tension or the surface tension of the domain wall 
that will dominate the dynamics when these configurations are formed after 
the chiral transition.

In the absence of anomaly the NA and abelian string configurations \cite{bals}
look superficially similar. But there are crucial differences. For in the 
core of the abelian string all the components of $\Phi$ vanish and so the full
chiral symmetry is restored. This is because the phase of all components of
$\Phi$ change by $2\pi n$ ($n$ being a non-zero integer) as one goes round an
abelian string obliging all these components to vanish at the core of the 
string. But for a NA string  one or more components of $\Phi$ need not change
in phase as the string is encircled, and hence need not vanish in the string
core. As a consequence the full chiral symmetry associated with a particular
linear combination of $T_0$ and $T_{N_f^2-1}$ for which the nonzero components
are singlets is restored. The other difference is as we have mentioned above, 
for $c \ne 0$, the elementary abelian string for $N_f$ flavours is connected 
to $N_f$ domain walls while the elementary NA strings (associated with
$M(\theta)$ and $M(\theta)^{-1}$) are connected to $1,2,..., N_f-1$ domain 
walls. (But note that $M^\prime(\theta)$ has two domain walls.)

\section{Metastable states in $U(3)\times U(3)$ linear sigma model}

As we have mentioned, in the chiral limit, anomaly breaks $U(1)_A$ to 
$Z_3$ in 3 flavour QCD. This leads to a set of discrete ground states related
by discrete transformations of $U(3)$. It is simple to find out the ground 
states in the chiral limit. Let us consider the effective potential in Eq.5,

\bea \label{L}
{\cal V}(\Phi) &=&
{\rm Tr}  \left(m^2 \, \Phi^{\dagger}\Phi \right) +
\lambda_{1} \left[ {\rm Tr}  \left( \Phi^{\dagger}
 \Phi  \right) \right]^{2} +
\lambda_{2} {\rm Tr}  \left( \Phi^{\dagger}
 \Phi  \right)^{2} \nonumber \\
&-& c \left[ {\rm Det} \left( \Phi \right) +
{\rm Det}  \left( \Phi^{\dagger} \right) \right]
- {\rm Tr} \left[H  (\Phi + \Phi^{\dagger})\right] \,\, .
\eea

\noindent In the chiral limit when $H=0$, one can choose the ground states to 
be of the form 
$\Phi_0 = \bar{\sigma_0} {T_0}/\sqrt{2}$. When the anomaly is also absent, 
$c=0$, and all $U(3)_A$ rotations of $\Phi_0$ cost no energy. For $c \ne 0$, 
only a few discrete axial rotations do not change the potential energy. Thus 
there are a few ground states. Some of the ground states are obtained by the 
discrete rotations $U \Phi_0 U $, where $U = exp(-i\theta_1/2)$, with 
$\theta_1 = {2\pi \over 3}, {4\pi \over 3}$. Also one can choose 
$U = exp(-i\theta_2 T/2)$, where $T$ is a linear combinations of $T_a$'s and 
$T_0$, for example $T_s$ in Eq.11. For $T = T_s$, $\theta_2 = \pi$ gives 
another ground state. All these ground 
states, except $\Phi_0$, will have non-zero $U(1)_A$ phase giving rise to 
$CP$ violating effects \cite{met}.

For the full potential (18), these states with $\theta_{1,2} \ne 0$ will be 
of higher energy. Still these states with differing values of $\theta_{1,2}$
are separated by energy barriers. When the quark mass is switched on, only one 
of these states becomes the true ground state. However, for small quark 
masses, the barrier between the ground state and other states may remain 
appreciable making them metastable. Because of their $CP$ violating
nature it is interesting to check if such metastable states can exist after
including the effects of the meson masses and their splittings. In the 
non-linear sigma model \cite{met} such states are known to appear for 
temperatures close to the deconfinement transition. Here we now show that 
such states exist also in the linear sigma model at zero temperature. Since 
these are due to effects of anomaly we expect that they will survive as long 
as anomaly effects are significant. 

%%%%%%%%%%%%%%%%%%%%%%%%%%%%%%%%%%%%%%%%%%%%%%%%%%%%%%%%%%%%%%%%
\begin{figure}[ht]
\begin{center}
%\vskip -1 in
\leavevmode
\epsfysize=7.0truecm \vbox{\epsfbox{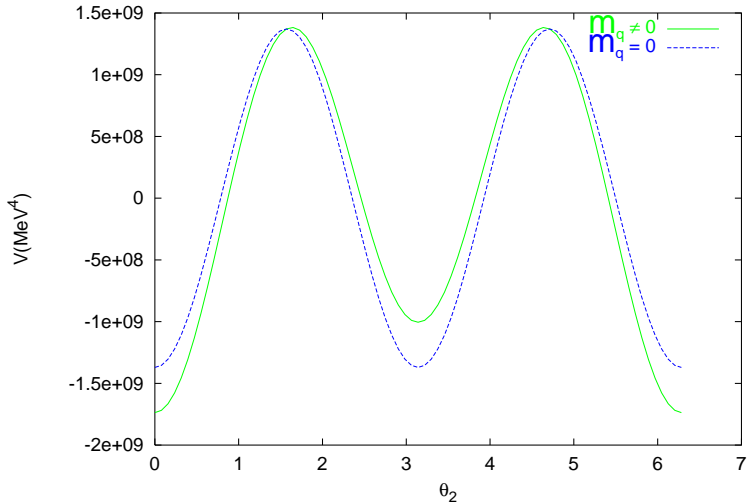}}
%\vskip -1in
\end{center}
\caption{ $\theta_2$ dependence of the potential energy for $m_q=0$
and $m_q \ne 0$.
}
\label{Fig.3}
\end{figure}
%%%%%%%%%%%%%%%%%%%%%%%%%%%%%%%%%%%%%%%%%%%%%%%%%%%%%%%%%%%%%%%%%%

In order to find metastable states,
we took the values of the parameters in Eq. (18) from \cite{jonathan} which
give an approximate description of meson masses and their splittings. The 
ground state is $\Phi_0 = \bar{\sigma_0} {T_0}/\sqrt{2} + \bar{\sigma_8} 
T_8$, where $\bar{\sigma_0} \sim 130 MeV$ and $\bar{\sigma_8} \sim -24 MeV$
while $H$ is $h_0{T_0}/\sqrt{2}+ h_8 T_8$, where $h_0 = (286.094 MeV)^3$ and 
$h_8 = -(310.960 MeV)^3$. Finally $\lambda_1 = 1.4$ and $\lambda_2 = 46.484$.
In order to see possible metastable states, we fix $\sigma_0$ and $\sigma_8$
and plot the $\theta_{1,2}$ dependence of the potential energy. In case
there is a local minimum for non-zero $\theta_{1,2}$, we then minimize
${\cal V}(\Phi)$ relaxing $\bar{\sigma_0}$ and $\bar{\sigma_8}$ fixing 
$\theta_{1,2}$. If the new $\bar{\sigma_0}$ and $\bar{\sigma_8}$ are close to 
their initial values, then it implies the existence of metastable states. We 
find no metastable states for non-zero $\theta_1$, 
so there exists no metastable
state with purely $U(1)_A$ phase. For $\theta_2$ there exists a metastable
state at $\theta_2=\pi$. In Fig. 3 we plot the $\theta_2$ dependence of the
potential energy for $m=0$ and $m_q \ne 0$. The figure clearly shows that 
there is a local minimum at $\theta_2 = \pi$ for $m_q \ne 0$. Fixing 
$\theta_2$ at $\pi$ we minimize ${\cal V}(\Phi)$ varying $\bar{\sigma_0}$ and 
$\bar{\sigma_8}$. The new values are $\bar{\sigma_0} \sim 120 MeV$ and 
$\bar{\sigma_8} = -28 MeV$ indeed very close to the initial values. 

\section*{Conclusions}

In summary, we have numerically found an approximate static solution of a
non-abelian string (domain wall) in the SSB of 
$U(N_f)\times U(N_f) \rightarrow U(N_f)$ in the linear sigma model. Also we 
have shown that metastable states exist in this model for realistic quark 
masses. Abelian and non-abelian extended regions in metastable vacua may 
form during the chiral transition ocurring in the early stages of heavy-ion 
collisions as well as in the early Universe. However, formation of strings 
and large metastable domains are complementary as formation of the one 
suppresses the formation of other. More the number of  uncorrelated domains,
more are the number of strings, with less probability for an extended domain 
in a metastable state.

\medskip

\noindent
{\bf Acknowledgments}

\medskip

This work was supported by DOE and NSF under contract numbers
DE - FG02 - 85ERR 40231 and INT - 9908763 and by BMFB (Germany) under grant
06 BI 902. We have benefited greatly from discussions with A. M. Srivastava.


\begin{thebibliography}{99}

\bibitem{vafa} C. Vafa and E. Witten, Nucl. Phys. {\bf B234}, 173
(1984); Commun. Math. Phys. {\bf 95}, 257 (1984).

\bibitem{hooft} G. 't Hooft, Phys. Rev. Lett. {\bf 37}, 8 (1976);
Phys. Rev. {\bf D 14}, 3432 (1976). 

\bibitem{berger} X. Zhang, T. Huang, and R. H. Brandenberger
Phys. Rev. {\bf D58} 027702, (1998); R. H. Brandenberger, and X. Zhang
Phys. Rev. {\bf D59} 081301, (1999).

\bibitem{bals} A. P. Balachandran and S. Digal, hep-ph/0108086 (Int. J.
Mod. Phys. {\bf A}, in press).

\bibitem{bal} A. P. Balachandran, G. Marmo, N. Mukunda, J. S. Nilsson,
E. C. G. Sudarshan, and F. Zaccaria, Phys. Rev. Lett {\bf 50}, 1553
(1983); {\it ibid} Phys, Rev. {\bf D29}, 2919 (1984);
{\it ibid} Phys. Rev. {\bf D29}, 2936 (1984); P. Nelson and A. Manohar,
Phys.Rev.Lett. {\bf 50}, 943 (1983); A. Abouelsaood, Phys.
Lett. {\bf B125}, 467 (1983) and references therein.

\bibitem{nab} G. 't Hooft, Nucl. Phys. {\bf B79},276 (1974); A. M. 
Polyakov, Pis'ma {\bf JETP 20}, 430 (1974).


\bibitem{met} D. Kharzeev, R. D. Pisarski and M. H.G. Tytgat,
Phys. Rev. Lett. {\bf 81} 512, (1998).

\bibitem{jonathan} J. T. Lenaghan and D. H. Rischke, Phys. Rev. 
{\bf D62},085008, (2000).

\bibitem{vilenkin} N. D. Mermin, Rev. Mod. Phys. {\bf 51}, 591 (1979);
A. Vilenkin and E. P. S. Shellard, `` Cosmic strings
and other topological defects'', (Cambridge University Press, Cambridge,
1994).

\bibitem{rose} C. Rosenzweig, J. Schechter, C. G. Trahern, 
Phys. Rev. {\bf D21}, 3388, (1980); A. Aurilia, Y. Takahashi, and 
P. K. Townsend, Phys. Lett. {\bf B95}, 265, (1980). 

\bibitem{pw} R. D. Pisarski and F. Wilczek, Phys. Rev. {\bf D29}, 338,
(1984).

\bibitem{alles} B. Alles, M. D'Elia, and A. Di Giacomo, Phys.Lett.
{\bf B483}, 139, (2000).


\end{thebibliography}
\end{document}